\documentclass[final,5p,times,twocolumn]{elsarticle}

\usepackage{amssymb}
\usepackage{bigfoot}
\usepackage{amsthm}
\usepackage{xcolor}
\usepackage{floatrow}
\usepackage{natbib}
\usepackage{balance}
\biboptions{sort&compress}
\usepackage{mathrsfs}
\usepackage{amsmath}
\usepackage{widetext}
\usepackage{hyperref}
\hypersetup{
  colorlinks   = true, 
  urlcolor     = blue, 
  linkcolor    = blue, 
  citecolor   = blue 
}

\journal{Combustion and Flame}

\begin{document}

\begin{frontmatter}


\title{Fractal dimension of premixed flames in multifractal turbulence}

\author{Amitesh Roy\corref{cor1}}
\ead{amiteshroy94@yahoo.in}
\cortext[cor1]{Corresponding Author}
\author{R I Sujith}
\address{Department of Aerospace Engineering, IIT Madras, Tamil Nadu - 600 036, India}

\begin{abstract}
{\color{black}In turbulent premixed flames, the fractal dimension of flame iso-surface is argued to be $\mathbb{D}=7/3$ for Damk\"ohler's large-scale limit ($Da>>1$) and $\mathbb{D}=8/3$ for Damk\"ohler's small-scale limit ($Da\sim\mathcal{O}(1)$) based on heuristic scaling arguments.} However, such scaling arguments do not consider the effect of the multifractal nature of turbulent kinetic energy dissipation on the flame surface. {\color{black} In this paper, we account for the effects of multifractal dissipation on the fractal dimension of low $Da$ turbulent premixed flames. We derive two corrections to the upper-limit of fractal dimension -- $\mathbb{D}=8/3+3/4(1-D_{1/4})$ and $\mathbb{D}=8/3+2/3(3-D_{1/3})$ -- which correspond to the change in the scalar flux and the total area of flame interface due to fluctuations in the inner cut-off scale arising from the intermittent nature of turbulent dissipation, respectively. \color{black}We further show that the second correction leads to an explicit dependence of the fractal dimension ($\mathbb{D}$) on the scaling exponent ($\xi$) of the velocity structure function through the relation: $\mathbb{D}=11/3+\xi$.} Thus, we explicitly quantify the effect of the multifractal nature of turbulence upon low $Da$ premixed combustion.
\end{abstract}


\end{frontmatter}


\section{Introduction}
\label{Introduction}
 The complex, non-Euclidean geometry of turbulent flows has led to the widespread use of fractal and related concepts to understand the phenomenology of turbulence \cite{mandelbrot1982fractal, sreenivasan1991fractals}. Most notably, scalar iso-surfaces in homogeneous, isotropic turbulence have been shown to have a fractal dimension of $\mathbb{D}=7/3$ using considerations of particle-pair diffusion and various scaling arguments \citep{hentschel1984relative, sreenivasan1989mixing}. These predictions have been tested and validated in controlled experiments \cite{sreenivasan1986fractal, sreenivasan1989mixing}. The most renowned example of this scaling law is the fractal dimension of clouds which was shown to be 2.34 ($\simeq7/3$) by Lovejoy \cite{lovejoy1982area}. Observed deviations in the predicted value of the fractal dimension have since then been attributed to the phenomenon of small-scale intermittency of turbulent flows \cite{sreenivasan1989mixing, meneveau1990interface}. Small-scale intermittency refers to the increasing non-Gaussian behaviour in dissipation quantities when one approaches scales close to Kolmogorov's length scale. The small-scale intermittency is tied to the multifractality in the dissipation field, leading to extreme-value fluctuations in dissipation quantities, localized non-uniformly throughout space \cite{sreenivasan1997phenomenology}.

A closely related problem to the statistical description of scalar surfaces in turbulence is the description of propagating interfaces in turbulence. Propagating interfaces are frequently encountered in the study of premixed combustion and are of practical importance. In the limit where flow time scales are much larger compared to combustion time scales (i.e., Damk\"ohler number, $Da=\tau_{\eta}/\tau_{\text{chem}}=(\eta/\delta_F)^2>>1$), we can unambiguously define a flame surface by considering an iso-temperature or iso-concentration surface \cite{borghi1985structure, peters2000turbulent}. In other words, internal flame processes are not affected by turbulent fluctuations, and the effect of turbulence is restricted only to wrinkles on the flame surface. Here, $\tau_\eta$ indicates the time scale with respect to a Kolmogorov scale ($\eta$) vortex and $\tau_{\text{chem}}$ the chemical processes associated with combustion and $\delta_F$ is the flame thickness.

Early studies considered the flame surface in the limit of $Da>>1$ as a passive interface with fractal dimension 7/3 similar to iso-surfaces in turbulence \cite{gouldin1987application, kerstein1988fractal}. However, Kerstein \cite{kerstein1988fractal} showed that the similarity in the estimate of fractal dimension to that of passive scalars in turbulence is only coincidental and the assumption of a fractal flame surface being passive is not physically plausible. It was shown that a dynamical balance between wrinkling due to turbulent convection and smoothing effects due to flame propagation is established at all length scales in the scaling range. The fractal dimension of 7/3 was recovered naturally by considering the balance between characteristic burning time governed by turbulent burning velocity and eddy turnover time. Later, Kerstein \cite{kerstein1991fractal} considered the effect of intermittency in turbulent kinetic energy dissipation and arrived at a corrected value of the fractal dimension, which was quite similar to the correction obtained by Meneveau and Sreenivasan \cite{meneveau1990interface}. However, the two approaches were quite different and implied the possibility of geometrical equivalence of different surfaces in turbulent flows.

{\color{black} Application of concepts of fractals in the description of turbulent flames, however, has remained confined to combustion in the corrugated flamelet regime ($Da>>1$). This has led to the development of various combustion models \cite{gouldin1987application, gouldin1989chemical, mantzaras1989fractals, gulder1991turbulent, gulder1995inner, north1990fractal, gulder2000flame}. In particular, researchers have been quite successful in obtaining closure for the unresolved flame area in large eddy simulation (LES) models \cite{lindstedt1999modeling, knikker2002priori, charlette2002power, fureby2005fractal, hawkes2012petascale}. However, the success has not translated nearly as well while modelling low $Da$ flames. For low $Da$ flames, a range of eddies has turnover times comparable to the reaction time scale. These eddies can penetrate and disrupt the preheat zone. Nonetheless, the fractal framework is still applicable as the burning rate of the flame depends on the flux of fuel across the iso-concentration surface which is being entrained by the eddies in the preheat zone, provided that there are no local extinctions. Thus, even for the case of low $Da$ flames, one can define the flame surface as an iso-surface of progress variable based on fuel mass fraction and extract statistical properties of the fractal iso-surface \cite{chatakonda2013fractal}. Thus, the fractal properties of low $Da$ flame surfaces are expected to differ from that of corrugated flamelets \cite{chatakonda2010modeling, chatakonda2013fractal}.}

For $Da\sim\mathcal{O}(1)$, Chatakonda et al. \cite{chatakonda2013fractal} showed that instead of a dynamical balance between wrinkling due to turbulence and turbulent flame propagation as considered by Kerstein \cite{kerstein1988fractal}, a balance between flame stretch {\color{black}and flame propagation} at the lowest length scales is necessary. A balance of the tangential flame strain and the effects of curvature and flame propagation leads to a modified inner cut-off scale known as the Obhukhov-Corrsin length scale ($\eta_{OC}$). The effect of this modified inner cut-off scale then leads to a modified prediction in the fractal dimension of low $Da$ ($Da\sim\mathcal{O}(1)$) flames which is equal to $\mathbb{D}=8/3$. {\color{black} This was verified through direct numerical simulation of hydrogen-air flame and thermonuclear flames in type 1a supernovae \cite{chatakonda2013fractal}.} 

The fractal dimension of $\mathbb{D}=8/3$ was also suggested by Mandelbrot \cite{mandelbrot1982fractal} for scalars in Gaussian random fields and Kolmogorov spectra. Later, through an altogether different approach, Constantin et al. \cite{constantin1991fractal} theoretically derived the limit $\mathbb{D}=8/3$ for non-reacting passive scalars and experimentally verified by considering scalar in the well-mixed region of a turbulent flow.

In this paper, we account for the effect of the multifractal nature of dissipation on the inner cut-off scale in low $Da$ flames. The fluctuations in the inner cut-off scale lead to a variation in the scalar flux and the total area of the flame iso-surface. We show, in the same vein as Sreenivasan et al. \cite{sreenivasan1989mixing} and Meneveau and Sreenivasan \cite{meneveau1990interface}, that the intermittent nature of dissipation leads to corrections in the fractal dimension of low $Da$ flames. {\color{black}We further show that the corrections in the fractal dimension due to change in the total flame area is intimately related to the scaling exponent ($\xi$) of the velocity structure-function.}

{\color{black} The paper is organized as follows. In \S\ref{Sec2}, we present the derivation of fractal dimension of low $Da$ flames based on Chatakonda et al. \cite{chatakonda2013fractal}. We also discuss the various physical implications associated with the limits of $\mathbb{D}=7/3$ and $\mathbb{D}=8/3$ for non-reacting and reacting turbulent flows. In \S\ref{Sec3.1-Correction due to scalar gradient}, we derive the correction to the fractal dimension as a result of variability in the total scalar flux due to the dependence of inner cut-off scale on the turbulent kinetic energy dissipation. In \S\ref{Sec3.2-Correction due to total area}, we derive the second correction to the fractal dimension by explicitly integrating over the flame iso-surface to account for the variation of the inner cut-off on the total flame area. In \S\ref{Sec4-Discussion}, we contextualize our results in cognizance of past results in reacting and non-reacting turbulent flows. We also discuss the connection of the correction to the fractal dimension with the scaling exponent ($\xi$) of velocity structure-function. Finally, we draw the conclusions of our study in \S\ref{Sec5 - Conclusions}.}

\section{Estimate of fractal dimension of low $Da$ flames}
\label{Sec2}
{\color{black} We define the flame surface as an iso-concentration surface of a progress variable based on fuel mass fraction. This allows us to define a flame surface even for low $Da$ flames where turbulent eddies penetrate the preheat zone. Even so, the net burning rate depends upon the flux of reactants across fuel iso-concentration surface. We assume that the flame is devoid of any local extinction.} We consider a flame surface, defined in this manner, propagating freely into a volume containing a combustible mixture. The flame surface divides the region of reactants and products. {\color{black} We focus on the wrinkles on the flame surface induced by turbulent fluctuations in the inertial range. This distinction is quite important as it removes any anisotropy due to the directional propagation of the flame surface and flow anisotropy due to effects such as mean shear, etc. Then, it follows that the fluctuations on the flame surface are locally isotropic, and the wrinkles follow dynamic self-similarity \footnote+{We emphasize that this assumption is only true at small scales and does not hold when the flame is viewed at the scale of the turbulent flame brush.}.} The outer cut-off is conveniently defined by the integral length scale of the flow ($\ell$). {\color{black}We further assume that the turbulence levels and Reynolds number are high enough such that outer cut-off ($\ell$) remains independent of the Reynolds number. In particular, the Taylor scale Reynolds number ($Re_\lambda$) should be greater than $50$ \cite{sreenivasan1984scaling, lindstedt1999modeling}}. We discuss the inner cut-off in more detail later on. The inner (outer) cut-off scale indicates the scale at which the measurement area no longer scales with an increase (decrease) in the resolution of the measurement. 

Under the assumptions made above, the true area of such a flame surface, $A_T$, depends on the measurement scale ($\epsilon_i$) through the power-law relation \cite{gouldin1987application}:
\begin{equation}
A_T(\epsilon_i) = A_0 \big(\epsilon_i/\ell\big)^{2-\mathbb{D}},
\label{Eq1-True_area_fractal_dependence}
\end{equation}
where $A_0$ is some normalizing area proportional to $\ell^2$ and $\mathbb{D}$ is the fractal dimension. 

For a stable and well-maintained flame surface in a turbulent flow field, turbulence {\color{black}induced} tangential flame strain at the lowest length scales are balanced by the effects of curvature and flame propagation \cite{hawkes2012petascale}. The tangential flame strain rate due to an eddy of size $\epsilon_i$ is $a_T=u_i^\prime/\epsilon_i$, where $u^\prime_i$ is the turbulent intensity at the scale $\epsilon_i$. Further, turbulent intensity can be written as $u^\prime_i\sim (\langle\varepsilon\rangle/\epsilon_i)^{1/3}$, where $\langle\varepsilon\rangle$ is the rate of turbulent kinetic energy dissipation averaged over the volume $\ell^3$. Thus, the tangential flame strain rate can be re-written as $a_T \sim (\langle\varepsilon\rangle \epsilon_i)^{1/3}/\epsilon_i$. 

The effect of flame propagation in balancing the tangential flame stretch is negligible for low $Da$ flames, and is well-supported by theory \cite{peters1999turbulent} and DNS results \cite{hawkes2005evaluation}. In such a case, the equilibrium on the flame surface is maintained by the effect of curvature alone \cite{hawkes2012petascale}. Curvature is quantified by the gradient of the surface normal $\nabla\cdot \textbf{N}$.  The balance between the curvature and the tangential flame strain is, thus, $\mathcal{D}\langle {\color{black}(\nabla\cdot\textbf{N})}^2\rangle_s\sim\langle a_T \rangle_s$, where $\mathcal{D}$ is the diffusivity and $\langle \rangle_s$ indicates average weighted by the surface area of the flame surface \cite{chatakonda2013fractal}. The balance then leads to
\begin{equation}
\langle (\nabla.\textbf{N})^2\rangle_s\sim \frac{1}{\epsilon_i^2} \frac{\epsilon_i^2}{\mathcal{D}}\frac{(\varepsilon \epsilon_i)^{1/3}}{\epsilon_i}.
\label{Eq2-Curvature_tangential_strain_balance}
\end{equation}
We can then define the Obhukhov-Corrsin length scale ($\eta_{OC}$) based on the balance above as \cite{chatakonda2013fractal}
\begin{equation}
\eta_{OC}\sim\epsilon_i\sim\big(\mathcal{D}^3/\langle\varepsilon\rangle\big)^{1/4}\sim Sc^{-3/4}\eta,
\label{Eq3-Obhukhov_Corrsin_length_scale}
\end{equation}
where the Schmidt number ($Sc$) is defined as the ratio of the kinematic viscosity ($\nu$) and the diffusivity ($\mathcal{D}$), i.e., $Sc=\nu/\mathcal{D}$. Here, $\eta$ is the Kolmogorov length scale and is related to the kinematic viscosity and mean turbulent kinetic energy dissipation as $\eta=(\nu^3/\langle\varepsilon\rangle)^{1/4}$.

We can assume that local isotropy is followed at the small-scales so that the length scale controlling the inner cut-off ($\epsilon_i\sim\eta_{OC}$) also controls the scalar gradient across the flame interface \cite{chatakonda2013fractal}. The scalar gradient is thus $\sim \delta c_{\eta_{OC}}/\eta_{OC}$, where $\delta c_{\eta_{OC}}$ is the difference in scalar concentration associated with length scale $\eta_{OC}$. For higher Reynolds number and in the limit of {\color{black}$Da\rightarrow\mathcal{O}(1)$}, the scalar difference can be related to the integral scale fluctuations ($c$) of the scalar through Kolmogorov's cascade argument such that $\delta c_{\eta_{OC}}\sim (\eta_{OC}/\ell)^{1/3}c$ \cite{chatakonda2013fractal}. 

The total scalar flux across the flame interface is, thus, proportional to the total area ($A_T$), the diffusivity ($\mathcal{D}$), and the scalar gradient ($\delta c_{\eta_{OC}}/\eta_{OC}$) -
\begin{align}
F&\sim A_T\mathcal{D}\big(\delta c_{\eta_{OC}}/\eta_{OC}\big)\\
&\sim A_0\mathcal{D} c  \big(\eta_{OC}/\ell\big)^{2-\mathbb{D}} \big(\eta_{OC}/\ell\big)^{1/3}\eta_{OC}^{-1}
\label{Eq4-flux_scaling_length_scale}
\end{align}
Using the definition of $\eta_{OC}$ from Eq. (\ref{Eq3-Obhukhov_Corrsin_length_scale}) and the scaling relation in the universal inertial subrange, $\eta/\ell\sim(Re)^{-3/4}$ where, $Re=u^\prime \ell/\nu$, we obtain:
\begin{equation}
F\sim A_0 c u^\prime(Sc)^{3/4(\mathbb{D}-8/3)}(Re)^{3/4(\mathbb{D}-8/3)}.
\label{Eq5-flux_scaling_Reynolds_number}
\end{equation}

Thus, we express the scalar flux in terms of only the Schmidt number and the Reynolds number. In the inertial subrange, the total scalar flux (mass, momentum, and concentration) is independent of the kinematic viscosity and Reynolds number. The independence of flux properties of the kinematic viscosity and energy inducing scale is referred to as the Reynolds number similarity. Thus, for $Re$ independence, the fractal dimension must be $\mathbb{D}=8/3=2.67$. This result was derived in Chatakonda et al. \cite{chatakonda2013fractal}.

{\color{black}A clarification regarding the physical mechanism leading to the two limits of fractal dimension -- $\mathbb{D}=7/3 $ and $8/3$ -- for both the reacting and non-reacting flow is in order. For non-reacting flow, $\mathbb{D}=7/3$ is observed in entraining layers of a passive scalar surface \citep{sreenivasan1989mixing}, and $\mathbb{D}=8/3$ for passive scalars in well-mixed regions \cite{constantin1991fractal}. The two limits pertain to the manner in which the characteristic difference between the scalar concentration changes ($\delta c_\eta$) as the inner cut-off ($\eta$) is reduced. For an iso-surfaces of an entraining outer superlayer, $\delta c_\eta$ is equal to the difference between the concentration levels between the ambient and the jet, leading to $\delta c_\eta/\eta\rightarrow\infty$ as $\eta$ diminishes. Thus, $\mathbb{D}=7/3$ corresponds to the large scale limit. Whereas, in the well-mixed  region, $\delta c_\eta/\eta\rightarrow 0$ as $\eta$ diminishes. In such a case, an iso-concentration surface appears fractal only above certain inner cut-off scale. Hence, $\mathbb{D}=8/3$ corresponds to the small-scale limit. Thus, the cross-over in the behaviour of the scalar concentration gradient leads to a cross-over in the value of $\mathbb{D}$ as one moves from the well-developed region (small-scale limit) to entraining region (large-scale limit) \cite{constantin1991fractal}.

In the case of a corrugated flamelet, Kerstein \cite{kerstein1988fractal} showed that one could arrive at $\mathbb{D}=7/3$ by considering the balance between convection and flame propagation in the fractal range. The dynamical similarity in the range of scales defined only by the outer cut-off naturally leads to $\mathbb{D}=7/3$, without the need for considerations for the scaling of the inner cut-off. Physically, in the flamelet regime, we notice that the characteristic scalar difference is determined by the sharp difference in the scalar concentration between the reactants and the products, and does not diminish when the inner cut-off is reduced. Thus, the limit of $\mathbb{D}=7/3$ corresponds to Damk\"ohler's large-scale limit \cite{damkohler1947effect}, and is analogous to the entrainment in non-reacting flows \cite{chatakonda2013fractal}. On the other hand, for low $Da$ flames, the limit of $\mathbb{D}=8/3$ is derived by considering a balance between flame stretch and propagation processes at the smallest scales. Alternatively, one can derive this limit by assuming that the inner cut-off in such flames depends on the Karlovitz number ($Ka = 1/Da$) and Damk\"ohler's small-scale limit applies. Simple algebra leads to the limit of $\mathbb{D}=8/3$ \cite{hawkes2012petascale}. Physically, in the thickened flame region, the characteristic scalar gradient diminishes to zero as the inner cut-off diminishes and the iso-concentration surface appears fractal only above the inner cut-off scale. Thus, the limit of $\mathbb{D}=8/3$ corresponds to Damk\"ohler's small-scale limit \cite{damkohler1947effect}, and is analogous to the fractal dimension of the scalar interface in the well-developed region \cite{constantin1991fractal}.}

\section{Corrections to the fractal dimension of the flame due to multifractal dissipation}
\label{Sec3-Correction}
\subsection{\color{black}Fluctuations in the inner cut-off scale and its effect on the scalar flux}
\label{Sec3.1-Correction due to scalar gradient}

{\color{black}So far}, we have assumed that the rate of turbulent kinetic energy dissipation remains uniform throughout the flow field, which allowed us to define a volume-averaged dissipation $\langle\varepsilon\rangle$ and subsequently, the averaged inner cut-off scales $\eta$ and $\eta_{OC}$. We notice from the definition that $\eta_{OC}$ and $\eta$ has a quarter power dependence on the dissipation rate. However, the energy dissipation of turbulent flows shows intermittent fluctuations in space. {\color{black} Fluctuations in dissipation will lead to fluctuations in the inner cut-off. Thus, we can obtain the correction by spatially averaging the flux after replacing averaged $\eta$ and $\eta_{OC}$ with their corresponding locally fluctuating value.} 

For an inhomogeneous dissipation field, the dissipation in a region of size $r^3$, embedded in volume $\ell^3$, can be approximated by the generalized power law \cite{sreenivasan1989mixing}:
\begin{equation}
\langle \varepsilon_r^q\rangle = \langle \varepsilon\rangle^q (r/\ell)^{(q-1)(D_q-1)},
\label{Eq6-Definition of local dissipation}
\end{equation}
where, $D_q$ indicates the generalized dimension of order $q$ and $\langle \varepsilon\rangle$ is the averaged dissipation in box of size $\ell^3$. We further note that for $q=2$, one obtains $\langle \varepsilon_r^2\rangle =\langle\varepsilon\rangle^2 (r/\ell)^{-\mu}$, where $\mu=1-D_2$ is the well-known intermittency exponent \cite{kolmogorov1962refinement}.

{\color{black}We are interested in measuring the flux across the interface allowing for the variability in quantities which are defined on the value of dissipation $\varepsilon_r$ such as $\eta$ and $\eta_{OC}$. Thus, instead of calculating the flux based on $\eta$ and $\eta_{OC}$ defined by average dissipation rate $\langle \varepsilon\rangle$, we measure the spatial average of flux based on $\eta$ and $\eta_{OC}$ defined by the unaveraged dissipation rate $\varepsilon_r$. This is essentially equivalent to replacing $\langle\varepsilon\rangle^{1/4}$ with $\langle \varepsilon^{1/4}\rangle$ in the defintion of $\eta$ and $\eta_{OC} $.} Thus, we replace $q=1/4$ in Eq. (\ref{Eq6-Definition of local dissipation}) and find that:
\begin{equation}
\langle \varepsilon^{1/4}\rangle \sim \langle \varepsilon_{\eta_{OC}}^{1/4}\rangle= \langle \varepsilon\rangle^{1/4} (\eta_{OC}/\ell)^{3/4(1-D_{1/4})}.
\label{Eq7-Local dissipation}
\end{equation}
Further, we can re-write Eq. (\ref{Eq3-Obhukhov_Corrsin_length_scale}) as:
\begin{equation}
\eta_{OC}\sim (\mathcal{D}^3/\varepsilon_{\eta_{OC}})^{1/4}.
\label{Eq8-InnerCutOff_Local_Dissipation}
\end{equation}
Substituting Eq. (\ref{Eq7-Local dissipation}) in Eq.(\ref{Eq8-InnerCutOff_Local_Dissipation}), we find:
\begin{align}
\frac{\eta_{OC}}{\ell}&\sim \frac{1}{\ell}\left(\frac{\mathcal{D}^3}{\varepsilon_{\eta_{OC}}}\right)^{1/4} \sim(Sc Re)^A,
\label{Eq9}
\end{align}
where, $A = -(3/4)/\big[1+3/4(1-D_{1/4})\big]$. Substituting Eq. (\ref{Eq9}) in Eq. (\ref{Eq3-Obhukhov_Corrsin_length_scale}) and carrying out the algebra, it is straightforward to see that the flux becomes:
\begin{equation}
F\sim A_0 c u^\prime(Sc Re)^B,\notag
\end{equation}
where, 
\begin{equation}
B = \frac{3/4\big[\mathbb{D}-8/3-3/4\big(1-D_{1/4}\big)\big]}{1+3/4(1-D_{1/4})} \notag.
\label{Eq10-Result1}
\end{equation}
Reynolds' similarity argument stipulates that $B=0$, from which we obtain:
\begin{equation}
\mathbb{D}=\frac{8}{3}+\frac{3}{4}\big(1-D_{1/4}\big).
\end{equation}

{\color{black}Equation (\ref{Eq10-Result1}) makes up our first result. We emphasize that the fluctuations in the dissipation rate are quite strong. However, the inner cut-off scale has a quarter power dependence on the dissipation rate, so that the strong variability in dissipation rate does not exactly translate to comparable fluctuations in the cut-off scale. Hence, we expect the corrections to be small.}

\subsection{\color{black}Fluctuations in the inner  cut-off scale and its effect on the total area of the interface}
\label{Sec3.2-Correction due to total area}

As mentioned above, the multifractal nature of dissipation leads to fluctuations in the inner cut-off $\eta_{OC}$ (Eq. \ref{Eq8-InnerCutOff_Local_Dissipation}). From Eq. (\ref{Eq1-True_area_fractal_dependence}) we observe that the total area depends on the inner cut-off. Thus, any fluctuations in the inner cut-off would lead to fluctuations in the total flame area. We show that this leads to the second correction to the fractal dimension of a stretched flame.
 
To find the corrections, we find the total flux by integrating boxes along the flame interface in the manner detailed in Meneveau and Sreenivasan \cite{meneveau1990interface}. We assume that the flame is contained in a domain of size $\ell^3$. We cover the entire domain in cubic boxes of size $\eta_{OC}^i$, which is the inner cut-off scale for our problem. Thus, the total flux is due to the sum of contributions of each of these boxes along the entire interface. The contributions of each box again depends on the area of the element ($(\eta_{OC}^i)^2$), the diffusivity ($\mathcal{D}$), and the scalar gradient ($\delta c_{\eta_{OC}}/\eta_{OC}^i$). Thus, the total flux after substituting the appropriate scalings associated with the gradient becomes:
\begin{align}
F\sim \sum_i \eta_{OC}^i \mathcal{D}\delta c^i_{\eta_{OC}}\sim c u^\prime \ell^2 (Sc Re)^{-1} \sum_i (\eta^i_{OC}/\ell)^{4/3}.
\label{Eq11-flux_along_interface}
\end{align}

Locally, for a domain of length scale $\eta_{OC}^i$, the dissipation depends on the local singularity strength $\alpha_i$ through the following relation:
\begin{equation}
\varepsilon_{\eta_{OC}}\sim\langle\varepsilon\rangle(\eta_{OC}^i/\ell)^{\alpha_i-3}.
\label{Eq12-dissipation_definition}
\end{equation}
The local singularity strength $\alpha_i$ is essentially the fractal dimension of the singularity associated with dissipation in a box of size $\eta_{OC}^i$, and changes when $\eta_{OC}^i$ changes. We can then express the inner cut-off $\eta_{OC}^i$ in terms of singularity exponent $\alpha_i$ alone by substituting Eq. (\ref{Eq12-dissipation_definition}) {\color{black}into} Eq. (\ref{Eq8-InnerCutOff_Local_Dissipation}) to obtain:
\begin{equation}
\eta^i_{OC}/\ell\sim\big(\eta/\ell\big)^{4/(\alpha_i+1)}(Sc)^{-3/(\alpha_i+1)}.
\label{Eq13-inner_cut-off_dependence_on_Kolmogorov}
\end{equation}

Now, we calculate the number of cubic boxes of size $\eta_{OC}^i$ having local singularity exponent $\alpha=\alpha_i$. The total number of boxes containing singularity exponent $\alpha_i$ in the box $\ell^3$ is defined by the scaling relation:
\begin{align}
N(\alpha_i)&\sim\left(\eta^i_{OC}/\ell\right)^{-f(\alpha_i)}.
\label{Eq14-falpha_relation}
\end{align}
Here, $f(\alpha_i)$ is the fractal dimension associated with counting the number of boxes of size $\eta_{OC}^i$ containing a given value of singularity exponent $\alpha_i$. Substituting Eq. (\ref{Eq13-inner_cut-off_dependence_on_Kolmogorov}) in Eq. (\ref{Eq14-falpha_relation}), we obtain
\begin{equation}
N(\alpha_i)\sim \left(\frac{\eta}{\ell}\right)^{-4f(\alpha_i)/(\alpha_i+1)}(Sc)^{3f(\alpha_i)/(\alpha_i+1)}.
\label{Eq15-falpha_relation-2}
\end{equation} 

In order to make further progress, we need to determine the total number of boxes of size $\eta^i_{OC}$ and singularity exponent $\alpha_i$ \textit{only along the flame interface}. Let $\mathcal{S}_1$ and $\mathcal{S}_2$ indicates the set containing the fractal flame element of dimension $\mathbb{D}$ and singularity $\alpha_i$ of dimension $f(\alpha_i)$ in 3-dimensional space $\mathbb{R}^3$, respectively. The co-dimension of $\mathcal{S}_1$ and $\mathcal{S}_2$ in $\mathbb{R}^3$ is thus $3-\mathbb{D}$ and $3-f(\alpha_i)$. The additive property of sets $\mathcal{S}_1$ and $\mathcal{S}_2$ stipulates: $\text{co-dim}(\mathcal{S}_1)+\text{co-dim}(\mathcal{S}_2)<\text{dim}(\mathbb{R}^d)$, which is the condition of intersection of sets $\mathcal{S}_1$ and $\mathcal{S}_2$ in a $d$-dimensional space \cite{mandelbrot1982fractal}. Physically, this implies that the flame surface intersects, or interacts, with the dissipation in the turbulent flow field. Thus, we have: $[3-f(\alpha_i)]+[3-\mathbb{D}]<3 \Rightarrow f(\alpha_i)+\mathbb{D}>3$. The dimension of the intersection of $\mathcal{S}_1$ and $\mathcal{S}_2$, i.e., the set of singularities $\alpha_i$ \textit{only along the} fractal flame interface,  follows from the condition of intersection as: $\mathscr{D}=f(\alpha_i)+\mathbb{D}-3$. Thus, the total number of boxes where $\alpha=\alpha_i$ along the interface is given by the dimension of the intersection of the sets $\mathcal{S}_1$ and $\mathcal{S}_2$ so that Eq. (\ref{Eq15-falpha_relation-2}) gets modified to
\begin{equation}
N(\alpha_i)\sim\bigg(\frac{\eta}{\ell}\bigg)^{-4\mathscr{D}/(\alpha_i+1)}(Sc)^{3\mathscr{D}/(\alpha_i+1)}.
\label{Eq16-Number_of_boxes_with_ai_along_interface}
\end{equation}

Note that we have cast the contribution of each box of size $\eta_{OC}^i$ to the total flux in terms of the distribution of singularity only along the interface. Thus, the summation in the contribution of all the individual boxes defined in Eq. (\ref{Eq11-flux_along_interface}) can be replaced with an integral over the entire spectrum of singularity exponent $\alpha$. The flux can then be calculated as
\begin{align}
F \sim c u^\prime \ell^2 (Sc Re)^{-1}\int N(\alpha)(\eta_{OC}(\alpha)/\ell)^{4/3} d\alpha.
\label{Eq17-total_flux_integral}
\end{align}

Substituting the expression for $N(\alpha)$ from Eq. (\ref{Eq16-Number_of_boxes_with_ai_along_interface}) and $\eta_{OC}$ from Eq. (\ref{Eq13-inner_cut-off_dependence_on_Kolmogorov}) in Eq. (\ref{Eq17-total_flux_integral}), we obtain
\begin{align}
F&\sim c u^\prime \ell^2 (Sc Re)^{-1} \int [(\eta/\ell) Sc^{-3/4}]^Cd\alpha,
\label{Eq18-total_flux_alpha_equation}
\end{align}
where, $C=-4(\mathscr{D}-4/3)/(\alpha+1)=-4[f(\alpha)-13/3+\mathbb{D}]/(\alpha+1)$. Equation (\ref{Eq18-total_flux_alpha_equation}) can be solved using the method of steepest descent in the limit of small $\eta<<\ell$. The saddle point is determined from $\partial C/\partial \alpha = 0$, which leads to:
\begin{equation}
\frac{df}{d\alpha}=\frac{f(\alpha)-13/3+\mathbb{D}}{\alpha+1}.
\label{Eq19-spectrum_q_dimension_criterion}
\end{equation}

Now, we know that $q=df/d\alpha$ which relates the order of the generalized dimension $q$ and the singularity specturm $f(\alpha)$. Further, the generalized dimension $D_q$ is related to the $f(\alpha)$ through the following relation \cite{halsey1986fractal}:
\begin{equation}
D_q=\frac{1}{(q-1)}[q\alpha-f(\alpha)].
\label{Eq20-Generalized_dimension_definition}
\end{equation}
Now, whenenver Eq. (\ref{Eq19-spectrum_q_dimension_criterion}) is satisified for a given $q$, we assign it the value $Q$ such that
\begin{equation}
\frac{df}{d\alpha}=\frac{f(\alpha)-13/3+\mathbb{D}}{\alpha+1}=Q=-C/4.
\label{Eq21-spectrum_q_dimension_criterion1}
\end{equation}
From here, we can work out the relation for $f(\alpha)$, which is
\begin{align}
f(\alpha) = Q(\alpha+1)+13/3 -\mathbb{D}.
\label{Eq22-falpha_Q}
\end{align}

The integral in Eq. (\ref{Eq18-total_flux_alpha_equation}) is evaluated at the saddle point where, in the power of the integrand, we substitute $C=-4Q$. In such a case, the total flux across the flame surface can be re-written, after substituting $\eta/\ell\sim Re^{-3/4}$, as:
\begin{align}
F&\sim c u^\prime \ell^2 (Sc Re)^{-1+3Q}.
\label{Eq23-total_flux_final_dependence}
\end{align}

Further, from Eqs. (\ref{Eq20-Generalized_dimension_definition}), (\ref{Eq21-spectrum_q_dimension_criterion1}) and (\ref{Eq22-falpha_Q}) we get
\begin{align}
D_Q &= \frac{Q\alpha -f(\alpha)}{Q-1},\notag
\end{align}
which after {\color{black}substituting} Eq. (\ref{Eq22-falpha_Q}) leads to 
\begin{align}
\mathbb{D}&=13/3+Q+(Q-1)D_Q.
\label{Eq24-fractal_dimension_generalized-dimension}
\end{align}

Now Reynolds similarity argument stipulates that the flux $F$ be independent of $Re$, which upon enforcing in Eq. (\ref{Eq23-total_flux_final_dependence}) we obtain $-1+3Q=0$, which yields $Q=1/3$. Consequently, we find from Eq. (\ref{Eq24-fractal_dimension_generalized-dimension}) the correction in the fractal dimension of the flame front as:
\begin{equation}
\mathbb{D}=\frac{8}{3}+\frac{2}{3}(3-D_{1/3}).
\label{Eq25-Fractal_dimension_correction}
\end{equation}

Equation (\ref{Eq25-Fractal_dimension_correction}) makes up the second result of our paper and is a direct result of the fluctuations in the area of low $Da$ flames due to small-scale intermittency.

\section{Discussion}
\label{Sec4-Discussion}
{\color{black} As mentioned earlier, Kerstein's \cite{kerstein1988fractal} result of $\mathbb{D}=7/3$ corresponds to the fractal dimension associated with Damk\"ohler's large-scale limit, comparable to the large-scale limit of $\mathbb{D}=7/3$ in a superlayer of entraining turbulent flow as found by Sreenivasan et al. \cite{sreenivasan1989mixing}. Later, Kerstein \cite{kerstein1991fractal}  derived the corrections to the fractal dimension due to dissipation-scale intermittency and observed the same correction as obtained for entraining turbulent flow by Meneveau and Sreenivasan \cite{meneveau1990interface}. On the other hand, the limit of $\mathbb{D}=8/3$ was derived by Chatakonda et al. \cite{chatakonda2013fractal} by accounting for Damk\"ohler's small-scale limit, which is similar to the dimension of the interface due to small-scale fluctuations inside the core of a turbulent jet \cite{constantin1991fractal}.

Equations (\ref{Eq10-Result1}) and (\ref{Eq25-Fractal_dimension_correction}), which quantify the effect of dissipation-scale intermittency on the fractal dimension of the interface associated with Damk\"ohler's small-scale limit, make up the main result of this study. We arrive at the two results through two key considerations. The first result is a consequence of spatially averaging the scalar flux across the interface after accounting for the variability in the inner cut-off due to dissipation-range intermittency, instead of calculating the average flux by considering an inner cut-off based on the average dissipation rate. The second result follows from accounting for the fluctuations in the total area of the interface as a result of fluctuations in the inner cut-off. 

The particular form in which the correction to the fractal dimension of low $Da$ flames has been expressed in Eq. (\ref{Eq10-Result1}) and (\ref{Eq25-Fractal_dimension_correction}) allows us to correlate the fractal dimension with the velocity structure function. We know that the moment of velocity difference ($\Delta \textbf{v}_r$) between two points separated by distance $r$ follows a power-law of the form:
\begin{equation}
\langle \Delta \textbf{v}_r^q\rangle\sim \langle [\textbf{v}(\textbf{x}+r)-\textbf{v}(\textbf{x})]^q\rangle\sim r^{\xi_p},
\end{equation}
where $\xi_q$ is the scaling exponent associated with order $q$ \cite{sreenivasan1997phenomenology}. For homogeneous isotropic turbulence, the scaling exponent is $\xi_q=q/3$, and follows directly follows from Kolmogorov's work \cite{kolmogorov1941local}. Intermittent nature of dissipation rate manifest in the scaling exponent deviating from $\xi_q=q/3$ at $p>3$. Following Meneveau and Sreenivasan \cite{meneveau1987multifractal}, we find that the scaling exponent is related to the generalized dimension as:
\begin{equation}
\xi_q = \bigg(\frac{q}{3}-1\bigg)D_{q/3}+1.
\end{equation}
Note that if $D_q=1$ for all $q$, dissipation field is space-filling and we get back $\xi_q=q/3$. For $q=1$, we find that  $\xi=1-2/3D_{1/3}$. Comparing this with Eq. (\ref{Eq25-Fractal_dimension_correction}), we observe that
\begin{equation}
\mathbb{D}=\frac{11}{3}+\xi.
\label{Eq26}
\end{equation}
Thus, the second correction in Eq. (\ref{Eq25-Fractal_dimension_correction}) shows that the fractal dimension of low $Da$ flames is also related to the scaling exponent of the velocity structure function.} 	

Experimentally measuring the multifractal spectrum of turbulent kinetic energy dissipation in turbulent reacting flows is a significant challenge as it involves the determination of two-point velocity correlations. The generalized dimension $D_Q$ which appears in Eqs. (\ref{Eq10-Result1}) and (\ref{Eq25-Fractal_dimension_correction}) can then only be measured from high fidelity DNS data. {\color{black} From the literature on turbulence, we find that the correction to $\mathbb{D}=7/3=2.33$ is due to $D_{1/3}$ and $D_{1/4}$ \cite{sreenivasan1989mixing, meneveau1990interface}. For the entraining turbulent superlayer, $D_{1/3}=0.96$ and $D_{1/4}=0.97$, leading to two estimates of $\mathbb{D}=2.36$ and 2.356, respectively. We see that the two estimates of the fractal dimension for an entraining non-reacting turbulent superlayer are numerically equivalent, even though each of these estimates arises from separate considerations. We believe that the corrections to the value of the fractal dimension of low $Da$ premixed turbulent flame could be relatively small. Nonetheless, the considerations made above are still important.} To the best of our knowledge, we did not find any study on turbulent combustion which have measured these dissipation quantities ($D_Q$, $f(\alpha)$ vs $\alpha$, etc.). So, we do not know the exact extent of variation in the estimate of $\mathbb{D}$. The determination of such quantities from DNS data appears worthwhile as it has the potential to lead to better closure models for low $Da$ flames used in LES.

\section{Conclusions}
\label{Sec5 - Conclusions}
In this study, we derive corrections to the fractal dimension of premixed flames in the limit of $Da\sim\mathcal{O}(1)$. {\color{black}We consider the flame surface to be defined by an iso-concentration surface based on progress variable of the fuel mass fraction. For low $Da$ flames, we consider the balance between kinematic viscosity and scalar diffusion and define the so-called Obhukov-Corrsin length scale. Considerations of the average flux across the flame surface leads to the upper limit of the fractal dimension of low $Da$ flames as $\mathbb{D}=8/3$. However, such a consideration does not account for the multifractal nature of turbulent kinetic energy dissipation, which leads to strong intermittent fluctuations in the dissipation field in space. Such fluctuations lead to significant variability in the inner cut-off, and consequently to the scalar flux and total area of the flame interface. Each of these considerations separately lead to two corrections in the estimate of the fractal dimension of low $Da$ flames (Eqs. \ref{Eq10-Result1} and \ref{Eq25-Fractal_dimension_correction}). We further show that the correction is intimately tied to the scaling exponent of the velocity structure-function (Eq. \ref{Eq26}).} Thus, we analytically quantify the effect of multifractal turbulent dissipation upon low $Da$ premixed combustion.

\section*{Acknowledgement}
We gratefully acknowledge the J. C. Bose fellowship (JCB/2018/000034/ SSC) from the Department of Science and Technology (DST), Government of India for the financial support. AR gratefully acknowledges the Ministry of Human Resource Development (MHRD), Government of India for funding PhD through Half-Time Research Assistantship (HTRA).




\bibliographystyle{elsarticle-num} 
\bibliography{references}

\end{document}